\documentclass[
11pt, 
english, 
singlespacing, 
headsepline, 
]{MastersDoctoralThesis} 

\usepackage[utf8]{inputenc} 
\usepackage[T1]{fontenc} 
\usepackage{amsmath}

\usepackage{palatino} 
\usepackage{hyperref}

\AtBeginDocument{}

\usepackage[autostyle=true]{csquotes} 
\usepackage{setspace}
\usepackage{amsthm}

\thesistitle{Human-Usable Password Schemas: \\Beyond Information-Theoretic Security} 
\supervisor{Prof. Manuel \textsc{Blum}} 
\cosupervisor{Prof. Santosh \textsc{Vempala}} 
\examiner{} 
\degree{SCS Honors Undergraduate Research Thesis} 
\author{Elan \textsc{Rosenfeld}} 
\addresses{} 

\subject{Computer Science Department} 
\university{Carnegie Mellon University} 
\department{Computer Science Department} 
\group{School of Computer Science} 
\faculty{School of Computer Science} 

\hypersetup{pdftitle=\ttitle} 
\hypersetup{pdfauthor=\authorname} 
\hypersetup{pdfkeywords=\keywordnames} 

\begin{document}

\frontmatter 

\pagestyle{plain} 


\begin{titlepage}
\begin{center}

\textsc{\LARGE \univname}\\[1.5cm] 
\textsc{\Large SCS Honors Undergraduate Research Thesis}\\[0.5cm] 

\HRule \\[0.4cm] 
{\huge \bfseries \ttitle}\\[0.4cm] 
\HRule \\[1.5cm] 
 
\begin{minipage}{0.4\textwidth}
\begin{flushleft} \large
\emph{Author:}\\
\authorname 
\end{flushleft}
\end{minipage}
\begin{minipage}{0.4\textwidth}
\begin{flushright} \large
\emph{Advisor:} \\
\supname \\

\emph{co-Advisor:} \\
\cosupname \\
\end{flushright}
\end{minipage}\\[3cm]
 
{\large April 29, 2016}\\[4cm] 
\includegraphics[height=6cm]{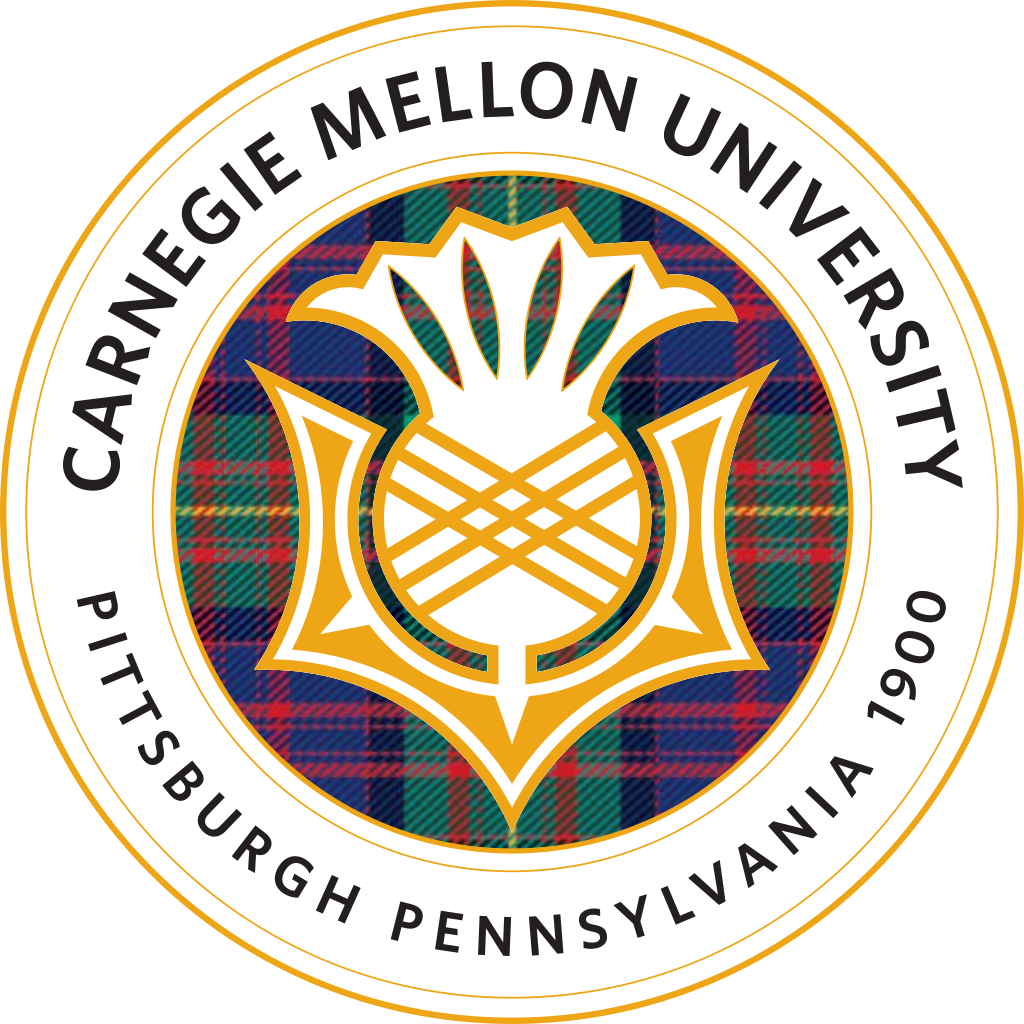} 
 
\end{center}
\end{titlepage}




 
 



\begin{abstract}
\addchaptertocentry{\abstractname} 

Password users frequently employ passwords that are too simple, or they just reuse passwords for multiple websites. A common complaint is that utilizing secure passwords is too difficult. One possible solution to this problem is to use a password schema. Password schemas are deterministic functions which map challenges (typically the website name) to responses (passwords). Previous work has been done on developing and analyzing publishable schemas, but these analyses have been information-theoretic, not complexity-theoretic; they consider an adversary with infinite computing power.

We perform an analysis with respect to adversaries having currently achievable computing capabilities, assessing the realistic practical security of such schemas. We prove for several specific schemas that a computer is no worse off than an infinite adversary and that it can successfully extract all information from leaked challenges and their respective responses, known as challenge-response pairs. We also show that any schema that hopes to be secure against adversaries with bounded computation should obscure information in a very specific way, by introducing many possible constraints with each challenge-response pair. These surprising results put the analyses of password schemas on a more solid and practical footing.

\end{abstract}


\begin{acknowledgements}
\addchaptertocentry{\acknowledgementname} 

I am thankful to Professor Santosh Vempala for his invaluable help and suggestions, including creating and helping to break several schemas.\\

Thanks to Dr. Jeremiah Blocki for his assistance with the constraint solver, as well as all his very constructive feedback on this paper. Thank you to Samira Samadi and Lisa Masserova for helpful discussion.\\

Thank you to my Abba for supporting and encouraging me through this process.\\

Finally, my deepest gratitude is to my advisor, Professor Manuel Blum. He gave me blunt criticism when I deserved it; he nonetheless later agreed to advise me, even when I hadn't proven myself. He encouraged me when I doubted myself and opened my eyes to the joys of research. I am deeply indebted to him for all this and much more.

\end{acknowledgements}


\tableofcontents 






\mainmatter 

\pagestyle{thesis} 



\chapter{Introduction} 
\label{Chapter1} 


\newcommand{\keyword}[1]{\textbf{#1}}
\newcommand{\tabhead}[1]{\textbf{#1}}
\newcommand{\code}[1]{\texttt{#1}}
\newcommand{\file}[1]{\texttt{\bfseries#1}}
\newcommand{\option}[1]{\texttt{\itshape#1}}


\begin{doublespace}
Password users frequently employ passwords that are too simple, or they just reuse passwords for multiple websites \cite{cranor, morris}. A common complaint is that utilizing secure passwords is too difficult. One possible solution to this problem is to use a password schema. Password schemas are deterministic functions which map \keyword{challenges} (typically the website name) to \keyword{responses} (passwords).

Previous work has been done on developing and analyzing publishable schemas. Some provide security even against adversaries who have seen a dozen random challenge-response pairs, but take more than an hour for a human to memorize, or more than a minute to generate responses \cite{hopper}. Others meet the human-usability requirement but are limited in security, or they only meet the usability requirement with the aid of a semi-trusted computer \cite{blum, blocki}. Importantly, each of these schemas has been almost exclusively analyzed through a theoretical lens.

When assessing the practical security of a schema, it becomes more beneficial to consider an adversary with bounded computation. As in \cite{blum}, we define for a given schema a metric known as \keyword{schema quality}, denoted $Q$ (occasionally we use $Q_X$ to refer to the schema quality of the schema $X$). We define a \keyword{challenge-response pair} as a word chosen from a given dictionary and its respective response; we define a \keyword{random challenge-response pair} as a \emph{randomly chosen} challenge and its response. A schema is said to have quality $Q$ if an information-theoretic adversary (i.e., an adversarial Turing machine with unbounded computational power) is able to correctly guess the response to the next challenge within ten attempts after seeing an average of $Q$-1 random challenge-response pairs. Given $Q$-1 challenge-response pairs, can a \emph{computationally bounded} adversary be expected to successfully guess the next correct response?

Certainly, there exist schemas for which this is not the case. Specifically, Blocki et al. \cite{blocki} showed a method of constructing a schema for which a program is believed to require far more than $Q$ examples to break, relying on the intractability solving random constraint satisfiability problems; Allison et al. \cite{allison} achieved the same, relying on the intractability of learning juntas. However, these schemas do not allow the user complete self-sufficiency: in the former, a user is presumed to have access to a semi-trusted computer. In the latter, the website is trusted to provide the same challenge on each visit (or else it would have to know the user's secret). Should an attacker compromise the website and become able to adaptively provide new challenges to the user, this schema is likely no longer secure.

Our goal then is to develop a schema which requires no involvement from any party except the user yet is powerful enough that a computer requires more than $Q$ examples on average to break it. These schemas are functions applied to an input that remains constant for each website; typically, we apply them to the domain name, though other possibilities exist---the challenge itself need not remain secure. Chapter 2 introduces the desirable traits of a password schema that meets our requirements.

With these desiderata established, Chapter 3 presents in detail a particular schema with a previously unknown upper bound for $Q$. We also introduce techniques for breaking these types of schemas and address the feasibility of constructing one that is secure against computationally bounded adversaries. Chapter 4 considers schema implementation from the user's perspective, introducing fundamental operations that a human can perform to transform a challenge into a response. Based on these operations, we introduce certain limitations on a schema which moderately constrain the time that a user has to respond to a challenge.

Using the same axiomatic operations, Chapter 5 presents an argument for why any schema that hopes to be secure must hide information in a very specific way. We then present and analyze one such schema, including the successes and failures of current attempts to break it.
\end{doublespace}

\chapter{What Makes a Good Password Schema}
\label{Chapter2}

\begin{doublespace}

A good password schema is one that meets a multitude of requirements for security and usability. The strength of the schema can be assessed by analyzing to what extent it meets each of them. We use the same criteria as \cite{blum}, restated briefly here for convenience.

\section{Desiderata}
A good password schema should be \textbf{publishable}, \textbf{human-usable}, \textbf{secure}, \textbf{self-rehearsing}, and \textbf{analyzable}.

\begin{enumerate}
\item \textbf{Publishable}

A schema is publishable if a detailed description of its implementation is publicly available; the security of the schema cannot rely on obscurity, except for the user's individual, secret key(s).

\item \textbf{Human-Usable}

The schema must be implementable in the user's head, without the use of additional instruments such as a calculator or pen and paper. We consider schemas with varying limits on the time a user should take to generate a response from any given challenge. We additionally require the bound of one hour on the total time over a user's life required to memorize and maintain the schema.

\item \textbf{Secure}

An adversary who knows the schema should have no better than random chance of being able to correctly guess responses to new challenges among those consistent with challenge-response pairs observed so far. Note that this definition does not limit the adversary to polynomial-time bounded computation---as a result, our definition of security is information-theoretic.


This definition comes from a game played between the user and an adversary: an impartial judge provides an adversary with a randomly chosen challenge from the dictionary, and the adversary gets 10 attempts to guess the correct response. If none of the guesses are correct, the judge provides the adversary with the response and repeats with a new challenge. This is repeated until the adversary correctly states the response to a new challenge. $Q$ is then the average number of challenges that the adversary needed to see in order to be able to guess the correct response, \emph{including the final one}. Note that $Q$ is dependent on the length of the challenge.

\item \textbf{Self-Rehearsing}

Use of the schema should result in frequent practice of every part of its implementation, such as applying functions over the whole domain. For example, if the schema involves a map on the domain of letters, then J, Q, and other uncommon letters might be mapped very infrequently. Steps must be taken to ensure that they are still practiced normally, so the user doesn't forget them.

\item \textbf{Analyzable}

The schema should be stated so explicitly that the instructions are able to be followed by a Turing machine. Note that we do allow for randomization over the choice of secret keys, but the domain over which the random choice is made and the method of randomization must be clearly defined and published.

\end{enumerate}

\section{Specific Definitions}

A few of our definitions or expectations are more specific than these general desiderata. Particularly, for our analysis of $Q$ we define the dictionary from which challenges are selected as the uniform distribution over all strings of a particular length using the twenty-six letters of the alphabet. Blum and Vempala \cite{blum} also consider a sample space of the most popular website names, but for a simpler analysis we stick with the uniform distribution over random strings.

In addition, we occasionally impose somewhat stricter limitations on response-time for the user. Blum and Vempala \cite{blum} originally imposed a maximum response-time of thirty seconds; we consider a schema which limits the user to one second per letter of the challenge. Since we primarily consider challenges of lengths between six and twelve characters, for the most part this limit more than halves the time available to the user.

\end{doublespace}
 
\chapter{Breaking Schemas with Bounded Computation}
\label{Chapter 3}

\begin{doublespace}
\section{Example Schema: Digit Schema 3 (DS3)}

We present one of the schemas analyzed in \cite{blum}: DS3, so titled because it utilizes a map from letters to digits. This schema is one of the simpler ones, with a high ratio of security to usability.

\subsection{DS3 Implementation}

A challenge $C$ consists of $L$ letters $A_1,...,A_L$ and the response consists of $L$ digits $b_1,...,b_L$. All addition is modulo 10.\\
The secret key consists of:

$f : [A-Z] \longrightarrow [0-9]$, a random map from the alphabet to digits

$g : [0-9] \longrightarrow [0-9]$, a random permutation on digits\\

\textbf{Let $DS3_{\langle f,g\rangle}(C)$ denote the response to $C$ under DS3 using secret maps $f$ and $g$.}\\

To determine $DS3_{\langle f,g\rangle}(C)$:\\

Output $b_1 = g(f(A_1)+f(A_L))$

For $i = 2$ to $L$:

\verb|  |Output $b_i = g(f(A_i) + b_{i-1})$\\

\subsection{Analysis}

In the guessing game that defines $Q$, each challenge-response pair that an adversary observes can provide some new information about the user's secret mapping. For an information-theoretic adversary, the simplest technique for breaking the schema would be to maintain all possible $\langle f,g\rangle$ pairs in a set $S$ and eliminate them as inconsistencies arise. Since all mappings are equally likely, the adversary is unable to distinguish between any $\langle f,g\rangle$ pairs that are so far consistent with the seen challenge-response pairs. Thus, the adversary is only guaranteed to correctly guess the response to the next challenge $C$ when $$\big|\{DS3_{f,g}(C) | \langle f,g\rangle \in S\}\big| \le 10$$ (though he could get lucky and guess it sooner).

We can see that the size of $S$ is reduced in expectation by at least a factor of 10
 with each seen challenge-response pair; if there are fewer than 10 unique responses by the remaining possible secret mappings, the adversary will guess the correct response and the game will end. Since all mappings are initially possible and the challenges are chosen at random, we expect the remaining possible mappings to distribute approximately evenly among the possible responses; this means that seeing the correct response should eliminate at least $\frac{9}{10}$ of the so-far consistent mappings, which implies that for any schema, $Q$ is roughly upper-bounded by the logarithm base 10 of the size of the secret key space.

For DS3, the key space for $f$ is $10^{26}$ and for $g$ is $9!$. Thus, the upper bound for $Q_{DS3}$ is roughly $\log_{10}(10^{26}\cdot 9!) \approx 31.56$. Simulating the above information-theoretic technique on random challenges of length 10, \cite{blum} gave an estimate for $Q_{DS3}$ of 6.91. However, this is still with the assumption of an infinite adversary---an actual computer obviously does not have the capacity to maintain all possible mappings in memory. This means that this value serves as only a lower bound for the security of the schema for a computationally bounded adversary. The question remains: can a computer break any schema with $Q$ pairs in expectation, or is the infinite computational power actually necessary to extract and utilize all the information contained in the challenge-response pairs? More generally, can a human being possibly do enough computation and obfuscation in a limited amount of time (thirty seconds, by the limit set in \cite{blum}) that a computer cannot extract all the available information?

As an alternative to holding all possible mappings in memory, suppose instead the adversary simply assigns an ordering to the secret keys---for the case of DS3, there is an obvious, easy ordering---and then iterates through them one at a time. At each possible key, the adversary tests if it is consistent with the observed challenge-response pairs. If so, he guesses the next response using that secret; if not, he eliminates it and moves on to the next one. This bypasses the memory issues imposed by the previous technique, but introduces a new problem: considering every possible key is intractable. So yes, a computationally bounded adversary \emph{can} extract all the information with limited memory, but whether or not it can be done in a reasonable amount of time remains to be seen. In order to build a solver that can hope to break a schema with $Q$ challenge-response pairs in a reasonable time frame, the problem must be approached more cleverly.

Note that in this context we are worried about \emph{actual running time}, rather than asymptotic complexity. Even if it requires a computer program exponential time to break a schema, it could still potentially succeed extremely quickly (i.e., a few seconds to a few minutes). This is because the the schema must be implemented by a human, which severely limits the size of the problem. So, instead of looking for a human-usable schema which requires any program exponential time to break, we are looking for one that takes a computer an \emph{extremely long time} to solve, at which point a user can simply change his secret key. We do not define an exact duration which we consider to be long enough; it should be considered acceptable if the user is willing and able to memorize a new secret key with the same frequency, to ensure security.

\section{Constraint Solving}

Because the schema is published, the adversary knows exactly how the secret key(s) were used to transform the challenge into the response. This means that each challenge-response pair shared with the adversary provides a set of constraints on the correct secret mapping, which can be used to eliminate inconsistent mappings without explicitly trying them. For example, a single challenge-response pair from DS3 results in $L$ constraints involving $f$ and $g$.

There is a wide range of methods for solving this problem, each one differently balancing the trade-off between flexibility and speed. Simple Gaussian elimination is incredibly fast; with this strength comes the fact that it only works for equalities. If the constraints on the secret includes inequalities, Gaussian elimination will not work. Note that we could choose to ignore those constraints in our solver. While this would certainly work, the fact that our solver is explicitly ignoring information contained in each challenge-response pair means that we will almost certainly require more than $Q$ pairs on average to break the schema.

If the constraints include inequalities, the next solution to consider is linear programming, where the objective to minimize is just a penalty function and incorrectly guessing the response is associated with an infinitely positive value. This is also quite fast, but has its own weakness: for linear constraints with a modulus, the objective function is non-convex. Since addition or multiplication with two or more digits takes humans quite a while without pen and paper, most human-usable schemas will use modular arithmetic. This means that for most human-usable schemas, linear programming is out.

This leaves two clear remaining possibilities: mixed integer linear programming (MILP) and constraint solving. MILP is linear programming with the ability to constrain certain variables to integers. With this method we can solve systems of linear inequalities with a modulus. MILP is quite slow---indeed, it is NP-Hard, as can be shown with a simple reduction from Vertex Cover\footnotemark[1]
---but it is still feasible since the number of constrained variables is limited by the amount of computation a human can do. Because the schema is implemented by a human, asymptotic complexity is not a concern.

\footnotetext[1]{https://en.wikipedia.org/wiki/Integer\_programming\#Proof\_of\_NP-hardness}

Constraint solving is more general than MILP. Constraint solvers actually utilize MILP solvers, but with some additional specific constraint types, such as specifying that certain sets of variables are all the same or all different. Constraint solvers can also have symbolic variables representing entire functions rather than single applications of a function; this unique feature means that as function applications are chained together, the number of variables in a constraint solver increases additively, as opposed to multiplicatively as in linear programming. Additionally, unlike MILP, constraint solving is focused on feasibility rather than optimization. This is actually more appropriate, because the only thing our schema solver needs to find is a secret key that is consistent with all previously seen challenge-response pairs. However, since finding a solution is equivalent to solving a MILP problem, it's more beneficial to analyze modern MILP limitations when considering solver capabilities.

Both of these methods are traditionally solved with the "Branch and Bound" method described in \cite{driebeek}---or a refined version of this technique known as "Branch and Cut"---and are therefore similar in complexity and runtime. As noted above, constraint solving allows for symbolic representation of function application constraints and the use of "higher-level" constraints which increases complexity but decreases the total number of constraints needed.

While \cite{blum} provided an estimate of 6.91 for $Q_{DS3}$ (for $L = 10$), to the best of our knowledge, no upper bound for a computationally bounded adversary has been previously defined. Using Microsoft Solver Foundation's constraint solver, we wrote a program to break DS3 according to the response-guessing game. How our solver cracks this schema is quite simple, because each new challenge-response pair reliably produces a new set of linear constraints. The constraint solver searches for possible values for the user's secret keys $f$ and $g$ that are consistent with previously seen examples and uses them to guess the next response. Following the rules of the game, if the program guesses correctly, the round is over. Otherwise, it is given the correct response---which is utilized by translating the challenge-response pair into $L=10$ new constraints and applying them to the system---and allowed to guess the response to the next one.

As the system has more and more constraints applied to it, the space it searches for a potential solution rapidly shrinks until it eventually correctly guesses the response. At this point we note the number of challenge-response pairs it saw and move on to the next round. This method of cracking this schema runs extremely quickly; a single round takes no more than a second, often much less. DS3 is an excellent example of how schemas can be moderately difficult for a human to implement and yet very simple for a computer to break quickly. Over several hundred-thousand experiments, we estimated an empirical upper bound for $Q_{DS3}$ of 6.89. Exactly calculating $Q$ is infeasible.

This result, that a computer can successfully extract all information included in a set of leaked challenge-response pairs, is the first instance of a true upper bound on the security of a schema for a computationally bounded adversary. While previous work was purely theoretical, for the first time we have concrete evidence that constraint solving can be used to break a schema in the smallest number of challenge-response pairs possible. This work further indicates that for any human-usable schema---until there is evidence otherwise---a pragmatic analysis should consider $Q$ as the actual number of challenge-response pairs that a real life adversary would require to break a user's secret keys. However, the \emph{amount of time} that an adversary would take to break said schema is still in question; this problem is addressed in Chapter 5.

\end{doublespace}
\chapter{Time-Bounded Schemas}
\label{Chapter 4}

\begin{doublespace}
To consider how a human can hide information from a computationally bounded adversary, we must first explicitly define the capabilities of a human when implementing a password schema. Blum and Vempala \cite{blum} gave a bound on time needed to apply the schema to a single challenge, but this bound does not consider specific actions to be carried out by a user. In order to get an idea of what kind of obfuscation a human-usable schema can achieve, we have to be able to break a schema into its individual operations and determine how they work together to hide information from an adversary.

\section{Operations}

We define a set of axiomatic operations, \textbf{OPS}, that we believe enumerate (!) all possible operations a human can perform with the letters of a challenge and any resulting values. To each of these operations we assign a time cost. These time costs were determined empirically and represent what we consider to be a reasonable lower bound on the amount of time the average user would require to complete an operation. More specifically, we derived these time costs by implementing a wide array of schemas with varying operations and then solving for the minimum individual operation times. This metric therefore relies on the assumption that the authors of this paper are not significantly slower than the average user. 

Using the estimated times for each operation, any analyzable schema can be broken down into its constituent operations and the amount of time to apply the schema to a particular challenge can be estimated. The individual elements of \textbf{OPS} are listed below, along with their respective time costs. Keep in mind that the time cost is for the \emph{average user}, and in cases where an exception might be made, it is listed.

\begin{enumerate}
\item \textbf{Perform} an arithmetic operation [$t_p = 0.35$ seconds]
\subitem -- This is for adding or multiplying two 1-digit numbers. It becomes more complex for larger numbers, but these can be broken down into combinations of 1-digit operations. This does \emph{not} include incrementing or decrementing a value that is being held in memory. If we are holding a single value in memory, incrementing or decrementing it costs $0.1$ seconds.
\subitem -- This operation relies on the user having memorized their times tables and being comfortable with addition. If the user has memorized larger addition or times tables then those are included.
\subitem -- If the operation is performed with a simple, intuitive modulus (i.e. 2, 5, or 10), the time cost does not change. If the modulus is more complicated, it counts as additional operations (see \textbf{search}, below). If the user has memorized other moduli or has an extremely fast method of calculating it (comparable to "keep the last digit" for mod 10), then these are included.
\item \textbf{Apply} a function (map) or permutation [$t_a = 0.3$ seconds]
\subitem -- This operation encompasses the application of any memorized function that accepts \emph{a single argument} (e.g., $f$ and $g$ in DS3). If a function accepts multiple arguments, a human will have to perform something akin to currying---considering one argument at a time---which will take longer.
\item \textbf{Iterate} to a value (context switch) [$t_i = 0.25$ seconds]
\subitem -- This operation is for any context switch. This includes considering a different value in memory, switching to a different function, and considering the next letter of the challenge.
\item \textbf{Output} a character [$t_{o_1} = 0.25$ seconds / $t_{o_2} = 0.15$ seconds]
\subitem -- This operation is for outputting any character, whether it was generated via a schema or is just meant to artificially lengthen the response (e.g., prepending the response with "aA@1" to meet a password requirement). Obviously a generated character takes longer to output; if the user knows the entire set of characters to be typed, it is much faster to consider the set of characters as a whole and type them together quickly. We found that outputting a character that was individually generated accrues a greater time cost $t_{o_1}$, while $t_{o_2}$ is the lower bound for the cost per letter of any other type of output.
\item \textbf{Classify/Compare} a character [$t_c = 0.1$ seconds]
\subitem -- This operation is for grouping letters or digits into simple, well-known categories, such as vowels/consonants or evens/odds. The categories must be known before learning the schema, commonly rehearsed, and limited to a small number of groups---this varies per user, but is probably at most three to four. For example, most users would not be able to classify a two-digit number as prime/composite within this time cost. If a schema required such a classification, this would fall under an \textbf{apply} operation\footnotemark[1].
\footnotetext[1]{\textbf{classify} is essentially the same as \textbf{apply}, but it takes less time because it's for already-known, often-practiced classifications (and it's limited to a small number of categories). This has the upside of costing less time, but the downside of being enumerable by a computer, so it can't be relied upon as the only method of obfuscation (see \textbf{Lemma 1}).}
\subitem -- Because our schema must be publishable and analyzable, any classification used in the schema must be explicitly stated (though the schema can list several options from which a user can choose at random). A schema that expects the user to choose their own method of classification is not analyzable and therefore does not fit our requirements. This requirement implies that the classifications that the user can choose from will be \emph{easily} enumerable by a computer; it is unreasonable that the average user could be expected to already know and frequently practice so many classifications that this is not true. We limit the number of possibilities to something on the order of $10^3$, which we consider to be a very generous limitation (our own brainstorming sessions provided closer to $10^2$ unique classifications).
\subitem -- This operation also includes simple comparisons, such as determining if a given number is larger than another number.
\item \textbf{Search} for a value [$t_s \ge 0.2$ seconds]
\subitem -- This operation, which is quite variable in terms of time cost, is for any time a user wishes to "fill in the blank" with a familiar value. For example, if the user is calculating 53 mod 7, he will likely \textbf{search} for a known multiple of 7 that is as large as possible without being greater than 53 (if the user must iterate through multiples of 7, this will obviously take longer than a single search operation). Having found 49, he will then subtract 49 from 53 to find the answer. In our tests, we found it common for users to also evaluate subtraction and division by performing a search operation (i.e., searching for the value $x$ such that $49+x = 53$).
\subitem -- We limit this operation to two inputs. If the calculation requires more than two inputs, it must be broken down into several chained applications.
\subitem -- Like classification, an analyzable schema would have to explicitly state how to use this operation, meaning if the user must choose at random from a set of possible ways to use the operation, all the choices will be enumerable by a computer---we use the same limitation of $10^3$.
\subitem -- We can further subdivide this operation into three categories:
\begin{itemize}
\item If none of the inputs are unknown to the adversary, then it doesn't hide any information; instead it is just a tool to help the user complete some other calculation such as division or subtraction.
\item If one of the inputs is unknown to the adversary, this operation is akin to an \textbf{apply} operation, with the distinction being that it requires more work to calculate, but is naturally self-rehearsing.
\item If both of the inputs are unknown to the adversary, this operation is similar to a \textbf{perform} operation, where the combination of the values is public but the values themselves are not.
\end{itemize}
\end{enumerate}

\subsection{LEMMA 1}

\textbf{Suppose a user must choose a combination of $x$ classify operations (chosen with replacement out of $N$ options) and $y$ search operations (chosen with replacement out of $M$ options) to be used to map a public input, with the restriction that $x\cdot t_c + y \cdot t_s \le t_a$. Then all such possible combinations are enumerable by a modern computer.}

According to our estimated time costs for the elements of \textbf{OPS}, this inequality limits us to $(x = 3, y = 0)$ or $(x = 1, y = 1)$. We specify this limitation because chaining together multiple applications of \textbf{classify} or \textbf{search} could theoretically be condensed into a single \textbf{apply} operation. So, if the time cost of this combination of operations is greater than the time cost of an \textbf{apply}, the user might as well just use an \textbf{apply} instead.

Since one of the inputs is public, this requires that any application of the \textbf{search} operation is of either the first or second subtype. We will exclusively consider those for which one of the inputs is private, as otherwise it is clear that the \textbf{search} operation will not increase the total number of possible combinations. 

By definition, both $N$ and $M$ are easily enumerable by a computer---we provided an upper bound of $10^3$. Given the limitations above, the total number of possible choices is equal to $\max(N^3, NM) \le 10^9$. With such a relatively small number of choices, a modern computer can easily simulate each combination on the public inputs. With this method, a program could maintain all consistent combinations, eliminating them as they are revealed to be impossible. \qedsymbol\\

Note that such an argument does not hold for the \textbf{apply} operation because the possible space of secret functions is larger than can reasonably be attacked with brute force by a bounded adversary; if this were not the case, then a computationally bounded adversary could break any schema with Q examples by enumerating all possibilities, as described above.

\section{Definitions}

We define a \keyword{one-pass schema} as any schema that consists of a single pass through the challenge, during which the user must use all the letters to produce the response.

By contrast, a \keyword{two-pass schema} is defined as any schema that can be divided into the following two phases: the first phase involves a single pass through the challenge, using as many or as few of the letters as desired. During this phase no elements of the response are outputted; instead all computations are done as a method of "seeding" the second phase. At the end of the first phase, the user will have some value to be used in the second phase---this could be something like a starting position, or something more complex such as a designation of which of several functions he should use. We refer to this value as the \textbf{seed}. The second phase proceeds as a one-pass schema, but the user can also use whatever seed he got from the first phase.

To consider the usefulness of the preliminary iteration through the challenge, we present the following way of thinking: consider the first pass as a process of traversing a directed acyclic graph (DAG), in which each node represents a set of intermediary values and each leaf represents a possible seed. Each time a user uses a letter of the challenge in this pass, he follows the appropriate edge to the next node, until he eventually arrives at the leaf which tells him the seed for the second pass through the challenge. Since no elements are outputted during the first phase, an adversary has no information about which leaf the user ended on; the idea being that if the user hides which seed he used for the second pass, much of the information leaked by challenge-response pairs will be obscured and difficult for the adversary to extract.

\section{Assumptions}

In addition to these definitions we'll assume that no elements in the response can be a direct mapping from an element of the challenge (i.e., the result of applying a map to a single element and then outputting it). This assumption makes it more likely that information in the secret mapping is properly obscured; otherwise a single challenge-response pair would leak too much information about the user's secret.

Here we point out that when considering the challenge, there is no way to use a particular letter to influence the result of either pass except by mapping it to some other value set---there are no already-defined operations on letters, and to create one would count as memorizing and applying a function. With this in mind, we'll add one final assumption: our schema cannot use a small number of \textbf{classify} and/or \textbf{search} operations as the sole method for converting any particular letter of the challenge into a value that can be used; by \textbf{Lemma 1} this would not have sufficient security. If the schema instead uses so many \textbf{classify} operations that the number of possibilities is too large to be "brute-forceable" (vulnerable to a brute-force attack), the amount of time such a combination would take for a user to evaluate would exceed that of just applying a map, so the time cost $t_a$ will serve as a lower-bound anyway.

These assumptions imply that in the second pass a map must be applied to all of the letters; we have to use all the letters, and applying a classification to any of the letters won't provide enough security, as explained above. Additionally, the map must be applied to each letter individually: given that the function must have only one input, the only alternative is to apply the function to multiple letters at a time. Memorizing a map on even just pairs of letters to anything else is infeasible; this function has a domain of size 676 (whose memorization would certainly exceed our stated bound of one hour over one's lifetime) and is also very clearly not self-rehearsing, as most letter pairs occur rarely, if at all.

\section{One- and Two-Pass Schema Analysis}

With these assumptions and time costs in hand, we can now consider the feasibility of creating a secure two-pass schema with a time bound of one second per letter for the user's response. 

\subsection{THEOREM 1}

\textbf{Let $S$ be a two-pass schema that can be precisely defined by some combination of the operations in \textbf{OPS}, where the time cost of those operations is limited to (on average) one second per letter. Define $S'$ as the one-pass schema which consists of just the second pass of $S$, where the seed that would have been determined in the first pass is unchanged but public. Let $N$ represent the number of possible seeds that could result from the first pass, and let $A$ be any computationally bounded adversary that can enumerate a set of size $N^{Q_{S}-1}$. If $A$ can solve $S'$ in $Q_{S'}$, then $A$ can solve $S$ in $Q_S$.}

This statement has one major assumption about the capabilities of the adversary, but it turns out that this assumption is actually not as hard to satisfy as it may appear. If we consider possible values for $N$, we can see that there are severe limitations on its maximum size. If $N$ designates a choice of function, it is limited to a very small number---probably no more than 3. Expecting a user to memorize more than that is unreasonable in the hour total memorization time limitation. Another possibility is that $N$ represents a starting position. As we'll show below, the limitation of one second per letter is so tight that performing just an additional 6 \textbf{iterate} operations is not possible, meaning $N\le 5$. Finally, $N$ could just be a value to be used in a \textbf{perform} operation; this would require it to be a single digit. So, $N$ generally appears to be restricted to around $10$; if it were much larger than this, using it would need several operations, which the time restriction really does not allow for.

Next we consider $Q_{S}$. As we will show, the time constraint ensures that $S$ is extremely limited in how it can transform a challenge into a response---so limited, in fact, that it behaves almost exactly like $DS3$. The fact that it can't really do anything more than map each letter and combine the results before outputting strongly implies that $Q_{S} \approx Q_{DS3}$. Considering the schemas defined in \cite{blum}, even those significantly more complicated than $DS3$ had schema qualities of no more than 12.

It is our belief that these two limitations are therefore realistic, which would imply that $N^{Q_{S}-1}$ is no greater than $10^{12}$. Given modern computational capabilities, this even allows for quite a bit of error in our estimates. We therefore claim that this requirement is not very difficult for an adversary to meet.

Finally, note that $S'$ gives strictly more information to the adversary than $S$, so $Q_{S'}$ is trivially less than or equal to $Q_{S}$.

\textbf{Proof of Theorem 1:}

Define $L$ as the length of the challenge and $n$ as the number of generated characters in the response. This means a user has $L$ seconds to respond with $n$ characters, plus whatever additional characters he adds on (e.g. for meeting website password requirements). We begin by considering the time costs of the second pass. In the second pass the user must \textbf{iterate} to each letter \emph{[$t_i\times L$ seconds]}, \textbf{apply} a map to each letter \emph{[$t_a\times L$ seconds]}, and ultimately he'll have to \textbf{output} $n$ generated characters for the response \emph{[$t_{o_2}\times n$ seconds]}. He still has to somehow combine the results of the mappings, otherwise a direct 1-to-1 mapping will appear in the response which violates our assumption. Since there are $L$ letters, there are two possibilities to combine these values. The user can \textbf{perform} an operation pairwise---this would result in a minimum of $\frac{L}{2}$ operations---but this will require him to \textbf{iterate} to each pair of values after he has mapped them \emph{[$(t_i + t_p) \times \frac{L}{2}$ seconds]}. Alternatively, he can keep the previous value in memory and update it with each new value as he comes to it; this is only a \textbf{perform} operation, but it must be done $L$ times \emph{[$t_p \times L$ seconds]}. This means the second pass, at the absolute minimum, will cost $(t_i+t_a+\min(\frac{t_i+t_p}{2}, t_p))L + t_{o_2}n = .85L+.25n$ seconds.


Clearly, the first pass can't use all the letters---to do so would require, at a minimum, iterating to each letter \emph{[$t_i\times L$ seconds]}. At this point we have already surpassed our limit of $L$ seconds for both passes, so using all the letters in the first pass is not a possibility.

Now we consider using some subset $m < L$ of the letters in the first pass. Because we're not using all of the letters, we also have to check at each letter if it should be included in the first pass. This is a \textbf{classify} operation and therefore costs the user an additional $t_c$ seconds per letter considered. So, for each of the $m$ included letters, the user must \textbf{iterate} to the letter, \textbf{classify} whether or not the letter is included, \textbf{map} the letter, and then \textbf{perform} an operation to follow an edge of the graph. This will cost the user $(t_i + t_c + t_a + t_p) \times m = m$ seconds to travel to a leaf of the DAG, with the generous assumption that he doesn't consider any letters that \emph{aren't} included.

Given that the total time cost of the schema applied to a challenge of length $L$ is limited to $L$ seconds, this gives us $.25n + m \le .15L$. Consider some possible values of $n$ and $m$. Suppose we only want to output one letter in the response for every two letters in the challenge---that is, $n = \frac{L}{2}$. Even with this reduction, this gives us $m \le 0.025L$. This implies that the first pass can use only $\frac{1}{40}$ of the letters of the challenge to traverse the graph, but this is too small of a fraction! Considering a limitation of 30 seconds to produce a response, the first pass couldn't even use one letter.

Reducing the fraction of letters that are output raises an additional complication: by doing the minimum of $\frac{L}{2}$ combination operations, the user generates $\frac{L}{2}$ values in the second pass. If he wants to output fewer characters than that, he will have to perform an additional \textbf{classify} operation to check whether a given value should be output. This gives the new inequality $.25n + m \le .1L$, further limiting the use of letters in the first pass. Clearly $n$ must be greater than 0; we consider a minimum of $n = \frac{L}{10}$. In this case, $m$ must be no more than $.075L$, meaning that less than one in thirteen of the challenge letters can be used for the first pass.\footnotemark[1]

\footnotetext[1]{It is here that we reference our previous claim regarding possible values for $N$: even with a maximum of $L = 30$, our user has at most $.075L = 2.25$ seconds for the first pass. If we want to use even one letter for the first pass, $m \ge 1$, which implies we would have an additional 1.25 seconds. This is enough time for at most 5 \textbf{iterate} operations.}

So, the first pass is limited to at most one or two letters of the challenge---even if $L$ is quite large and $n$ is small---meaning the number of possible paths through the DAG describing the first pass would be very small and easily brute-forceable by a computer. Since it's brute-forceable, the adversary $A$ could behave as if it were unbounded, simulating all possible paths.

Since each seed can take on one of $N$ values, after $Q_{S}\!-\!1$ challenge-response pairs there are $N^{Q_{S}-1}$ possible combinations of seeds. As this is enumerable, $A$ can simply solve $S'$ with each possible combination and eliminate inconsistencies---this is equivalent to making the seeds publicly known. This means that in expectation, once $A$ sees $Q_{S}\!\!-\!\!1$ pairs, it will be able to determine the correct seed (or, at the very least, a seed that will allow it to correctly guess the next response) for each seen challenge. Using those seeds, $A$ can work backwards, brute-forcing possible paths that were taken in the first pass, which will uncover enough information about the user's secret to determine the seed for the next challenge. Since $A$ has already seen at least $Q_{S'}\!-\!1$ challenge-response pairs, it will be able to break $S'$ and give the correct response to the $Q_{S}^{th}$ challenge, which means $S$ has also been broken in $Q_S$. \qedsymbol\\

So, a two-pass schema (limited to one second per letter) is computationally no harder to break than a one-pass schema. What if we drop the first pass and spend the entire time allotment on just a one-pass schema? The same assumptions and limitations would apply to the single pass of this schema as do to the second pass of a two-pass schema. Certainly, we could just output $\frac{L}{2}$ letters in the response and be done, though this schema functions almost exactly the same as DS3, which we know can be broken in $Q$. What other options remain? If we decide not to output all $\frac{L}{2}$ generated values, we have to \textbf{classify} each letter as explained above, which gives us $.25n \le .1L$. This means that $n \le \frac{2L}{5}$; with a challenge length of 15, the user would only produce a response of length 6.

In spite of this apparently negative result, this technique is actually a step in the right direction for possibly creating a password schema that a computer cannot break in a reasonable amount of time with $Q$ challenge-response pairs. As we discuss in the next chapter, a schema that hopes to achieve this level of practical security must obfuscate information in a specific way, by hiding which set of constraints correctly encodes the information of each challenge-response pair.

\end{doublespace} 
\chapter{Feasibility of Future Schemas}
\label{Chapter 5}

\begin{doublespace}
\section{What Can Computers Break?}

With the establishment of modern constraint solving techniques and definitions of and limitations on a human's computational capability, we are ready to more generally consider the possibility of schemas that---in expectation---a computationally bounded adversary requires more than $Q$ challenge-response pairs to break. We begin by considering how each of the operations a human can perform could hide information from an adversary and what distinguishes the capabilities of an infinite adversary from that of a computer.

Of the six operations that make up \textbf{OPS}, it is immediately apparent that \textbf{iterate} and \textbf{output} don't actually hide any information. These two operations are exclusively to allow the user to carry out the rest of the schema and are irrelevant to security. The remaining four operations each hide information in a specific way such that an adversary knows exactly how to define constraints given a challenge-response pair and knowledge of the operation's use.

For any publishable password schema, a challenge-response pair provides the adversary with one or more sets of possible constraints on the user's secret key. For a given challenge length $L$, we define the \keyword{expansion factor} of a schema as the expected number of possible sets of constraints encoded by a single challenge-response pair, denoted $F_L$. This value represents how quickly we expect the number of possible systems of constraints to grow as a function of the number of seen challenge-response pairs. Specifically, after seeing $n$ challenge-response pairs, an adversary would be expected to have to consider on the order of $F_L^n$ different combinations of sets of constraints. 

We now define two exhaustive and disjoint categories of password schemas: \keyword{direct} and \keyword{indirect}. For a given challenge length $L$, a password schema is said to be \keyword{direct} if each challenge-response pair corresponds to exactly one set of constraints with no ambiguity (i.e., for all direct password schemas, $F_L = 1$). DS3 is one such schema; every element of the response $b_i$ is equal to $g(f(A_i) + b_{i-1})$ (except for $b_1 = g(f(A_1) + f(A_L))$), which means that each challenge-response pair gives the adversary a single set of exactly $L$ constraints. These $L$ constraints convey the information contained in the challenge-response pair, and they can be applied to a solver to help it guess the correct mappings. In general, the constraints that are derived from a challenge-response pair of a direct schema include \emph{all of the information} included in that challenge-response pair. Since the schema is both direct and analyzable, the resulting constraints will give the constraint solver exactly as much information as it gives to an information-theoretic adversary.

\keyword{Indirect} password schemas are those for which $F_L > 1$. In other words, for an indirect schema, each challenge-response pair could encode one of several sets of constraints---they hide from the adversary which set of constraints is correct. When an adversary sees a challenge-response pair from an indirect password schema, each of the possible sets of constraints potentially encodes all information in that pair, but only one is correct.

It's important to note that indirect schemas have wildly varying expansion factors. If $F_L$ is reasonably small (less than 10), a constraint solver can handle the incomplete knowledge with an "OR" constraint, which requires that at least one of several sets of possible constraints is true. If it's much larger than that, solving the schema becomes a lot more complicated; given that the schemas we've considered tend to have a $Q$ around 7, a constraint solver using "OR" constraints would have to store and work with approximately $F_L^7$ constraint \emph{sets}, with each set containing several constraints. We found this to be well beyond the bounds of computation with today's available machines: in our experiments, when $F_L > 10$, the solver couldn't find a solution even with several days of runtime.

\subsection{CONJECTURE 1}

\textbf{Let $S$ be any direct, human-usable schema, limited to 30 seconds for generating any single response. Then a modern desktop computer, using a state-of-the-art constraint solver and/or MILP solver, can break $S$ with $Q_S$ challenge-response pairs in expectation in no more than 24 hours.}

Constraint solvers from as early as 1970 could solve problems with thousands of integer variables and tens of thousands of unconstrained linear variables. Practically, these algorithms were bounded by running time, which increased as a factor of several inputs---primarily the number of integer variables, since the tree that must be searched grows exponentially with this factor \cite{ibm}. Modern constraint solvers tend to only limit the number of variables by the amount of memory available to the system and can handle tens of millions of constraints. Complex MILP problems with thousands of integer variables and constraints are solvable by today's constraint solvers in less than 20 hours; with closer to 500 integer variables, these problems can even be solved in as little as a few minutes \cite{zhou, jain, timpe}.

A constraint solver---at the absolute maximum---has to define one variable for each element of the domain of each function that the user memorizes, as well as for any other hidden variables the user defines. Given the limit on memorization time, these are clearly less than a hundred. Additionally, new variables are created for modular arithmetic---for every modular equality the system must create an additional integer variable which represents the multiple of the modulus that is the difference between the value before and after the modulus operation. Even if every single \textbf{perform} operation carried out by the user creates a new variable, this would mean the limit of thirty seconds per response implies at most \textasciitilde175 integer variables, and that's an extremely conservative estimate.

These calculations imply that any direct schema can be solved extremely quickly and with expected $Q$ challenge-response pairs by a standard modern desktop. This would mean that any schema hoping to be unbreakable by a computer in $Q$ must be indirect.

Unfortunately, we are not yet able to prove this. There do exist certain especially hard MILP problems with a few hundred integer variables that have yet to be solved by state-of-the-art MILP and constraint solvers such as CPLEX and GLPK \cite{cplex}. While it seems quite likely that the human-usability and time limitations on these schemas would ensure that they are not quite so difficult, further analysis is needed to conclusively show this.

\subsection{Methods to Break an Indirect Schema}

So how hard is it to break an indirect schema? To consider this, we visualize a set of challenge-response pairs as a tree, where each challenge-response pair is a node, and from each node extends one edge for each possible set of constraints that it could imply. All nodes in a particular level of the tree are the same challenge-response pair, but each one represents a different combination of sets of constraints from the previous challenge-response pairs in the tree. In order to solve the schema, an adversary must traverse the tree, pruning branches as they reveal themselves to be inconsistent. Each time he reaches a new depth, he is given ten more guesses, which he makes based on any so-far consistent paths that he has found.

When an adversary is playing the guessing game, any secret that is not contradictory to seen challenge-response pairs is possibly the correct secret. On any particular branch of the tree, as long as there remains at least one secret key that is consistent, the adversary has no way of knowing whether or not that combination of sets of constraints is correct. Thus, the only way the adversary can definitively eliminate a branch of the tree is by showing that there is no secret mapping that is consistent with the assumed constraint sets on that branch.

The best method we've found to break a schema with $F_L > 10$ is to perform a simple depth-first search, assuming a particular set of constraints for each challenge-response pair and proceeding as if it is correct. When our algorithm finds that there is no solution that abides by the assumed constraints, it backtracks to the last decision point and tries a different assumption. This is the technique we employed to break our most promising schema, \textbf{Skip-To-My-Lou}\footnotemark[1], described below. Note that it may be possible to eliminate a branch from consideration without actually solving it, but rather by using some other heuristic that shows its inconsistency. Such an algorithm would be significantly faster than our approach.

\footnotetext[1]{Thanks to Santosh Vempala for suggesting this schema}

\section{Example Indirect Schema: Skip-To-My-Lou (STML)}

\subsection{STML Implementation}

A challenge $C$ consists of $L$ letters $A_1,...,A_L$ and the response consists of $m$ digits $b_1,...,b_m$, $0 \le m \le L$.\\
The secret key consists of:

$f : [A-Z] \longrightarrow [0-9]$, a random map from the alphabet to digits\\

\textbf{Let $STML_f(C)$ denote the response to $C$ under STML using secret map $f$}.\\

To determine $STML_f(C)$:
\newpage

Initialize $s = 0$, $j = 0$

For $i = 1$ to $L$:

\verb|  |$s = (s + f(A_i))$ mod 10

\verb|  |if $s \ge 5$:\\
\verb|        |Output $b_j = s$\\
\verb|        |$j = j + 1$\\

This schema is called "Skip-to-my-Lou" because its implementation consists of outputting the running total (mod 10) of the map applied to the challenge but skipping over values that are less than 5.

\subsection{Analysis}

STML is an excellent example of a simple, human-usable, indirect schema. For a challenge of length $L$ the response will be of length $m = \frac{L}{2}$ in expectation. For an adversary seeing a response of length $m$, he will have to guess which $m$ of the $L$ elements of the challenge resulted in an outputted digit and which did not (i.e., at which indices was the running total less than 5). This means there are $\binom{L}{m}$ possible constraints that an adversary could apply after seeing one challenge-response pair. It follows that for a given $L$, STML has an expansion factor of $\displaystyle\frac{1}{2^{L}}\displaystyle\sum_{i=0}^L \binom{L}{i}^2$. $F_5\approx 7.88$, which means that a solver can handle STML with $L=5$ with an "OR" constraint. The beauty of this schema is that as $L$ grows, the amount of work that the user does increases linearly, while the number of constraint sets the computer must consider grows exponentially. This exciting result implies that STML has the potential to be unbreakable by a computationally bounded adversary in $Q_{STML}$. Figure 5.1 (below) displays the astonishing growth rate of STML's expansion factor as a function of $L$.

With a very feasible challenge length of 10, $$F_{10} = \displaystyle\frac{1}{2^{10}} \displaystyle\sum_{i=0}^{10} \binom{10}{i}^2 \approx 180$$ which well exceeds a constraint solver's capabilities using an "OR" constraint. Even better, a randomly chosen set of constraints will likely not result in an unsolvable constraint system until the adversary has seen quite a few challenge-response pairs. Adding a new set of constraints, even an incorrect one, will often still allow the solver to find a secret that is consistent with all seen challenge-response pairs, but most of the time it will incorrectly guess the next response. This means that our solver has to travel several levels deep into the tree before it is able to label any constraints "impossible" and prune that branch (in our experiments for $L = 10$, the vast majority of eliminations---over 95\%---were made at a depth of at least 4). The result is a schema that forces an adversary to attempt to solve the system of constraints tens of millions of times. Where before, with DS3, our program could solve the system and break the schema in a fraction of a second, now it must do so for every possible combination of constraints until it is lucky enough to correctly guess the next response.

{\centering
\begin{figure}
\makebox[440pt]{\includegraphics[height=3.5in]{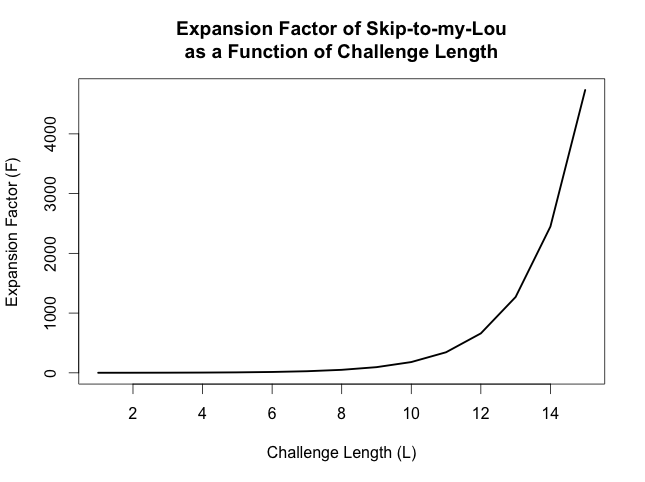}}
\caption{$F_L$ vs. L for Skip-to-my-Lou}
\end{figure}
}

The process of solving a system of constraints using a constraint solver can be very effective, but it also can be much slower than just coding the logic directly. Applying and removing constraints from a bulky constraint solver millions of times is a massive slowdown, so instead we constructed a new solver, specific to the constraints of STML. This new solver uses much more memory, but eliminates possible combinations of constraints at a significantly faster rate. Our results from attempting to break STML for varying values of $L$ are shown in Table 5.1 below. While exact values for $Q$ cannot feasibly be calculated, we can achieve a rough estimate as follows:

Define $m$ as the size of the alphabet of a schema's challenge; the size of the key space is therefore $10^m$. For each challenge-response pair that the adversary sees, in expectation half of the constraints---that is, $\frac{L}{2}$ of them---will be equality constraints, and the other $\frac{L}{2}$ will be inequalities. An equality constraint is expected to reduce the consistent key space by a factor of 10, while an inequality only eliminates half of the possible values, so it reduces the space by a factor of 2. However, there are roughly $\dbinom{L}{\frac{L}{2}} < 2^L$ possible sets of constraints. This means that each challenge-response pair can be expected to reduce the size of the key space by a factor of no less than $\frac{10^{\frac{L}{2}} \cdot 2^{\frac{L}{2}}}{2^L} = 5^{\frac{L}{2}}$. So, after seeing $k$ challenge-response pairs, the key space should be reduced by about a factor of $5^{\frac{kL}{2}}$. Setting this equal to the size of the key space and solving gives us $k = \frac{2\log_2(10^m)}{L\log_2(5)}$.

This $k$ roughly approximates the maximum number of challenges that an adversary would need to solve STML; after seeing $k$ challenge-response pairs, the adversary should be able to eliminate all but one of the possible secret keys. This means that $\lceil k\rceil$ is an approximate upper-bound on $Q$; an adversary is likely to be able to guess the correct response without knowing the exact key because, for any given challenge, a large portion of the key is expected to be irrelevant when calculating the response. Specifically, as $L$ increases, the portion of the secret key that is relevant to any given challenge grows, meaning $\lceil k\rceil$ will be better at estimating $Q$. Table 5.1 includes values for $\lceil k\rceil$ where $m = 26$.\\

\begin{center}
\begin{tabular}{| c | c | c | c | c |}
\hline
$L$ & $F_L$ & $Q$ & $\lceil k\rceil$ & Most challenge-response pairs\\
(challenge length)& (expansion factor) & & & in a single round\\
\hline
3 & 2.5 & 8.45 & 25 & 22\\
\hline
4 & 4.38 & 9.53 & 19 & 18\\
\hline
5 & 7.88 & 9.46 & 15 & 15\\
\hline
10 & 180.43 & 7.87 & 8 & 7\\
\hline
20 & 131,460.7 & (?) & 4 & N/A\\
\hline
\end{tabular}\\

Table 5.1: $L$, $F_L$, $Q$, $\lceil k\rceil$, and the most pairs needed in a single round\\
in experiments with our STML solver with $m = 26$
\end{center}

Up to $L = 10$, our solver can find a solution to an unknown STML key with expected $Q_{STML}$ challenge-response pairs in a very small amount of time (the longest round took about 5 minutes). However, due to the exponential growth of the tree that our solution must search, a user can make it quite a bit harder for our algorithm with a relatively small increase in his own computation. For example, if the user simply doubles $L$, he will now have to spend a little over 20 seconds per response, rather than 10. By contrast, our algorithm would have to brute-force attack a space that is approximately at least half a million times larger, probably much more than that. Indeed, even for a large fraction of rounds with $L = 15$, our solver couldn't break the schema even once despite several days of runtime. It is immediately apparent that a better technique is necessary for the adversary to keep up with this growth. 

\section{Future Considerations}

The existence of a simple, human-usable schema with such a large expansion factor appears to indicate one of two things: either STML is unbreakable by a modern computer in $Q$, or there is some way to eliminate branches of the constraint tree via a heuristic that does not require traversing deep into the tree. We wrote two separate solvers to break STML: the first iterates through the possible constraint combinations for each individual challenge-response pair and then adds them to a constraint system for a solver to use. This technique utilizes the branch and bound method, but updating the constraint solver object is much too slow to be done so many times. The second solver uses simple enumeration and elimination of possible mappings and successfully broke STML for $L=10$. Unfortunately, this method uses quite a bit of memory
, and also it has no possible way of looking ahead to future branches. As such, it performs worse for larger challenges.

We imagine that a joint solver that takes advantage of each of these two approaches' strengths might be able to do significantly better. That is, our hypothetical solver would quickly eliminate possibilities like the lightweight solver, but might use some heuristic defined by the constraint solver to determine which branches it should consider first, with the hopes of increasing its chances of finding the solution quickly or being able to prune closer to the root of the tree.

At the same time, it is quite likely that there are human-usable schemas with significantly larger expansion factors than STML. Because the tree of possible constraint combinations grows exponentially in the size of the expansion factor, a schema with a larger $F_L$---or perhaps just one that hides more information from a constraint solver---could be vastly more difficult to solve. As we work towards creating a faster solver, we hope to continue to push it to its limits with more and more challenging schemas; the end goal being either a human-usable password schema that a computer cannot break with only $Q$ examples, or a proof that none exists.

\end{doublespace} 

 




\nocite{*}
\bibliography{IEEEabrv,example}

\begin{thebibliography}{10}
\providecommand{\url}[1]{#1}
\csname url@samestyle\endcsname
\providecommand{\newblock}{\relax}
\providecommand{\bibinfo}[2]{#2}
\providecommand{\BIBentrySTDinterwordspacing}{\spaceskip=0pt\relax}
\providecommand{\BIBentryALTinterwordstretchfactor}{4}
\providecommand{\BIBentryALTinterwordspacing}{\spaceskip=\fontdimen2\font plus
\BIBentryALTinterwordstretchfactor\fontdimen3\font minus
  \fontdimen4\font\relax}
\providecommand{\BIBforeignlanguage}[2]{{%
\expandafter\ifx\csname l@#1\endcsname\relax
\typeout{** WARNING: IEEEtran.bst: No hyphenation pattern has been}%
\typeout{** loaded for the language `#1'. Using the pattern for}%
\typeout{** the default language instead.}%
\else
\language=\csname l@#1\endcsname
\fi
#2}}
\providecommand{\BIBdecl}{\relax}
\BIBdecl

\bibitem{cranor}
L.~F. Cranor, ``What's wrong with your pa\$\$w0rd?'' (2014, March), [Video
  file]. Retrieved from
  https://www.ted.com/talks/\\lorrie\_faith\_cranor\_what\_s\_wrong\_with\_your\_pa\_w0rd?language=en.

\bibitem{morris}
\BIBentryALTinterwordspacing
R.~Morris and K.~Thompson, ``Password security: A case history,'' \emph{Commun.
  ACM}, vol.~22, no.~11, pp. 594--597, Nov. 1979. [Online]. Available:
  \url{http://doi.acm.org/10.1145/359168.359172}
\BIBentrySTDinterwordspacing

\bibitem{hopper}
\BIBentryALTinterwordspacing
N.~J. Hopper and M.~Blum, \emph{Advances in Cryptology --- ASIACRYPT 2001: 7th
  International Conference on the Theory and Application of Cryptology and
  Information Security Gold Coast, Australia, December 9--13, 2001
  Proceedings}.\hskip 1em plus 0.5em minus 0.4em\relax Berlin, Heidelberg:
  Springer Berlin Heidelberg, 2001, ch. Secure Human Identification Protocols,
  pp. 52--66. [Online]. Available:
  \url{http://dx.doi.org/10.1007/3-540-45682-1_4}
\BIBentrySTDinterwordspacing

\bibitem{blum}
\BIBentryALTinterwordspacing
M.~Blum and S.~Vempala, ``Publishable humanly usable secure password creation
  schemas,'' in \emph{Proceedings of the Third AAAI Conference on Human
  Computation and Crowdsourcing}, 2015. [Online]. Available:
  \url{https://www.aaai.org/ocs/index.php/HCOMP/HCOMP15/paper/viewFile/11587/11430}
\BIBentrySTDinterwordspacing

\bibitem{blocki}
\BIBentryALTinterwordspacing
J.~Blocki, M.~Blum, and A.~Datta, ``Human computable passwords,'' \emph{CoRR},
  vol. abs/1404.0024, 2014. [Online]. Available:
  \url{http://arxiv.org/abs/1404.0024}
\BIBentrySTDinterwordspacing

\bibitem{allison}
S.~Allison, J.~Blocki, and M.~Blum, ``A secure human-computable authentication
  scheme from the k-junta problem,'' from personal communication.

\bibitem{driebeek}
\BIBentryALTinterwordspacing
N.~J. Driebeek, ``An algorithm for the solution of mixed integer programming
  problems,'' \emph{Management Science}, vol.~12, no.~7, pp. 576--587, 1966.
  [Online]. Available: \url{http://dx.doi.org/10.1287/mnsc.12.7.576}
\BIBentrySTDinterwordspacing

\bibitem{ibm}
\BIBentryALTinterwordspacing
M.~Benichou, J.~M. Gauthier, P.~Girodet, G.~Hentges, G.~Ribiere, and
  O.~Vincent, ``Experiments in mixed-integer linear programming,''
  \emph{Mathematical Programming}, vol.~1, no.~1, pp. 76--94, December 1971.
  [Online]. Available: \url{http://dx.doi.org/10.1007/BF01584074}
\BIBentrySTDinterwordspacing

\bibitem{zhou}
J.~Zhou, ``{Computational Experiments for Local Search Algorithms for Binary
  and Mixed Integer Optimization},'' Master's thesis, Massachusetts Institute
  of Technology, September 2010.

\bibitem{jain}
\BIBentryALTinterwordspacing
V.~Jain and I.~E. Grossmann, ``Algorithms for hybrid milp/cp models for a class
  of optimization problems,'' \emph{INFORMS J. on Computing}, vol.~13, no.~4,
  pp. 258--276, Sep. 2001. [Online]. Available:
  \url{http://dx.doi.org/10.1287/ijoc.13.4.258.9733}
\BIBentrySTDinterwordspacing

\bibitem{timpe}
C.~Timpe, ``Solving planning and scheduling problems with combined integer and
  constraint programming,'' \emph{Operations Research-Spektrum}, vol.~24, pp.
  431--448, October 2002.

\bibitem{cplex}
``{IBM} guidelines for estimating cplex memory requirements based on problem
  size,'' \url{https://www-01.ibm.com/support/docview.wss?uid=swg21399933},
  accessed: 2016-04-26.

\end{thebibliography}


\end{document}